\documentclass[accepted]{uai2023} 
\usepackage{mathtools} 
\usepackage{booktabs} 
\usepackage{tikz} 
\usepackage[american]{babel}

\title{An $\alpha$-Stable Paretian approach to modelling highly speculative assets and  cryptocurrencies}

\author[1]{\href{mailto:<gnf18@mails.tsinghua.edu.cn>?Subject=Your UAI 2023 paper}{Taurai Muvunza}{} }
\date{}
\affil[1]{%
    Tsinghua-UC Berkeley Shenzhen Insitute\\
    Shenzhen International Graduate School\\
    Tsinghua University\\
    Shenzhen, Guangdong, China
}
  
\begin{document}
\maketitle

\begin{abstract}
  We investigate the behaviour of cryptocurrencies using data for bitcoin, ethereum and ripple which account for over 70\% of the cryptocurrency market. We demonstrate that \(\alpha\)-stable distribution is an appropriately sufficient model for highly speculative cryptocurrencies which outperforms other heavy tailed distributions that are used in financial econometrics. We find that the maximum likelihood method proposed by DuMouchel (1971) produces estimates that fit the cryptocurrency return data much better than the quantile based approach of McCulloch (1986) and sample characteristic method by Koutrouvelis (1980). The empirical results show that the leptokurtic feature presented in cryptocurrency return data  can be captured by an \(\alpha\)-stable distribution. The findings highlight that \(\alpha\)-stable distribution is not only parsimonious with its four free parameters but also a creative model that is close to reality. This paper covers early reports and literature on cryptocurrencies and stable distributions. 
\end{abstract}

\section{Introduction}\label{sec:intro}
Bitcoin, ethereum and ripple are members of a family of cryptocurrencies whose return behaviour exhibit features that are inconsistent with traditional commodities and stocks.The return distribution of cryptocurrencies is characterised by skewness, a higher peak and heavy tails in contrast to those of normal distribution. Buttressing the distribution of price changes for any asset is vital for risk analysis and portfolio management. In this paper we model the cryptocurrencies with \(\alpha\)-stable distribution, and compare the goodness of fit to other distributions that are commonly used in financial econometrics. Our results show that \(\alpha\)-stable is the best distribution for the return data of cryptocurrencies. This paper is structured as follows; the first section reviews literature on cryptocurrencies, section 2 describes the data, section 3 introduces the empirical model used and covers literature on application of stable distributions. We review parameter estimation techniques in section 4 before discussing results and goodness of fit test in section 5. We conclude in section 6. 

Cryptocurrencies have received a lot of attention in mainstream media but their return behaviour has not been fully examined. While structural breaks in returns and volatility of bitcoin are frequent (Thies and Molnar, 2018) this paper proposes a distribution that can be utilised in the cryptocurrency market given any return series of the currencies under study. Bitcoin has characteristics that are similar to gold and the dollar (see Dyhrberg, 2016), one extreme being the pure store of value and the other extreme being pure medium of exchange. As a store of value such as gold, cryptocurrencies should not generate cash flow but rather retain their value (Cheah and Fry, 2015). However, bitcoin, for instance, is characterised by high volatility which makes it possible to earn enormous returns. The weak correlation between cryptocurrencies and equity markets (Bouri and Molnar, 2017) enhances the attractiveness of the currencies. When pulled together with traditional assets, bitcoin increases the value of a portfolio (Trimborn and Härdle, 2017) and can further serve as a hedge, safe haven or diversifier for other equity indices (see Bouri, 2017). Much of the academic literature on cryptocurrencies focuses on price volatility and legal aspects, and a more comprehensive analysis related to the behaviour of returns is necessary. 

In his work "Two Concepts of Money", Goodhart (1998) stressed that severe transaction costs in barter could lead to an evolution in search for cost minimization procedures within a private sector system, with which the government has no control at all. Cryptocurrencies have emerged as part of this evolution. Bitcoin was first introduced by pseudonym Satoshi Nakamoto, who argued that traditional financial institutions that are established on trust-based models have high transaction costs since they cannot avoid meditating disputes as third parties. Nakamoto (2008) further proposed that an electronic system based on cryptographic proof instead of trust is one of the ways third party costs can be avoided. Bitcoin, Etherium, Ripple, amongst other cryptocurrencies are an innovation that simplifies payment without the need for a third party. Bitcoin is believed to have been first minted on January 4th in 2009, its first payment occurred on January 11th and the software was publicly released as open source on the 15th of the same month and from there onwards, anyone with required technical skills could participate. 

With a market capitalization that reached 314 billion dollars in December 2017, bitcoin is an independent currency that has become popular among investors, consumers and retailers. The growing popularity of the use and acceptance of cryptocurrencies suggest that they can become an alternative currency in the future. European Central Bank (2012) pointed that an increase of electronic commerce particularly digital goods, growing access to and use of internet, higher degree of anonymity and lower transactions will precipitate the growth of digital currencies in the future. Cryptocurrencies are a form of digital currencies that are different from deposits. Dwyer (2014) clarifies this difference in that deposits are represented by a bank account balance at an institution while digital currency is viewed as storage of value which can be transferred without the intermediation of a financial institution. In the absence of this intermediation, digital currencies must not allow users to spend their balances more than once. To avoid double spending, Bitcoin uses peer-to-peer networks and open source software which generate computational proof of the chronological order of transactions, secured by a system of verifiable nodes, (Nakamoto 2008). Bariviera et al (2017) note that a distributed ledger which came with the invention of bitcoin is a key innovation in decentralizing and democratizing the currency. Distributed ledger is a consensus of replicated and synchronized digital data shared across the world by a group of peers who share responsibility for maintaining the ledger, (Deloitte, 2016).

Researchers have expressed diverging views on whether bitcoin is a currency or not. Yermack (2013) shows that with an exchange rate volatility that is higher than commonly used currencies, bitcoin exhibit zero correlation with other currencies and concluded that bitcoin does not behave as a currency. The rise in price of bitcoin from 4.951 cents on the first day of trading in July 2010 (Yermack, 2013) to its highest peak of 18 737.60 US dollars (coinmarketcap.com) in December 2017 shows that the criticism does not seem to thwart the demand for bitcoin. The total number of cryptocurrencies has exceeded 5000 in over 20 000 markets, commanding a total market capitalization of 270 billion dollars of which over 60\% is dominated by bitcoin (coinmarketcap.com, 2020).

Since inception, cryptocurrencies have become widely acceptable as a medium of exchange across the world. Hankin, (2017) records the first use of bitcoin to have been a pizza bought for 10 000 bitcoins in 2010, and afterwards, internet reports propagate the use of bitcoin to purchasing of illegal drugs, raising concern over its anonymity. Despite its popularity, the road to fame for bitcoin has been marred by challenges stemming from money laundering, drug dealing, fraud and security concerns. With approximately one billion dollars worth of bitcoin in circulation in 2013, the U.S Senate Committee set up a hearing to look into bitcoin, fearing that the system was a vehicle for money laundering and drug dealing. The hearing occurred after FBI shut down Silk Road, a website which sold illegal goods and drugs in bitcoin, (Dwyer, 2014). BBC (2013) reports that the currency trebled after news that the committee was told cryptocurrencies were legitimate financial services comparable to other online payment systems. Another major blow hit bitcoin in February 2014 when the Tokyo based Mt. Gox, the first and largest bitcoin exchange trading platform which handled over 70\% of all bitcoin transactions worldwide immediately suspended all transactions. According to Popper and Abrams (2014), Mt. Gox filed for bankruptcy citing "a weakness in our system" referring to what Hern (2014) called a loophole in bitcoin system that was exploited by hackers to get over 800 000 free bitcoins which accounted for 6\% of the total bitcoins at the time.

In academic literature, cryptocurrencies have not been fully examined. Dwyer (2014) focused on the price and returns of bitcoin and concluded that bitcoin is 10 times more volatile than stocks. Cheah et al (2015) found that bitcoin exhibits speculative bubbles and concluded that the fundamental value of bitcoin is zero. Blau (2017) argued that speculative trading does not explain bitcoin‘s price volatility. Bouri et al (2017) used a dynamic conditional correlation model to examine whether Bitcoin acts as a safe haven or hedge for stocks, bonds, oil and gold; and the empirical results conclude that Bitcoin is a poor hedge and can be used to eliminate idiosyncratic risk only. Dyhrberg (2015) demonstrated that bitcoin behaves like a currency and it has many similarities with gold and the dollar, one extreme being the pure store of value and the other extreme being pure medium of exchange. This gives bitcoin more advantages over other currencies as it can be used as an asset for portfolio management in the financial market. Urquhart (2017) examined the efficiency of bitcoin and found that bitcoin market is still inefficient but moving towards an efficient market. Bradvold et al (2015) concluded that bitcoin exchanges have significant contributions to bitcoin‘s price discovery due to information sharing. Ali et al (2014) showed that digital currencies such as bitcoin do not pose a material risk in the United Kingdom because they are only exploited by a few people. Trimborn, Li and Härdle (2017) showed that cryptocurrencies add value to a portfolio and using Markowitz optimization framework, they demonstrate that their approach can increase return of a portfolio while lowering volatility. Briere, Oosterlinck and Szafarz (2015) included bitcoin to a portfolio of traditional assets and arrived at the same conclusion. Our work focuses on modelling the returns of crytpocurrencies. In this paper we model the top three cryptocurrencies that account for over 70\% of the cryptocurrency market capitalisation return data using stable distribution. Understanding the distribution of returns is critical in evaluating risk and managing a portfolio of assets.

\section{Data}

To calculate returns, we used the first difference of cryptocurrency's log close price from 1 Dec 2011 to 31 Dec 2017,  7 August 2015 to 20 April 2018, 5 August 2013 to 20 April 2018 for bitcoin, ethereum and ripple, respectively. We use daily data from Yahoo Finance and it was stationary at first difference of log close price. Table \ref{tab:summary} shows summary statistics of bitcoin return data and we note that, for instance, ripple data has a kurtosis of 22.325 which is by far more than the kurtosis that is fit for a normal distribution.
\begin{table}[ht!]
    \centering
    \begin{tabular}{llllllllr} 
    \hline
       Crypto & Mean &Std.dev& Skewness & Kurtosis& Min & Max & J-B test & Obs \\ \hline
       BTC & 3.655E-03 & 0.0642& 4.888 & 160.27 & -0.849 & 1.474 & 2.261E6 & 2 187 \\
       ETH & 0.489E-02 &0.083 &-1.214 &20.901 & -0.916 & 0.3383 & 5.926 & 971\\ 
       XRP & 0.286E-02 & 0.121 & 0.747 & 22.325 & -0.997 & 1.028 & 5.939 & 1165\\ 
\hline
\end{tabular}
\caption{Summary statistics of cryptocurrency data}
\label{tab:summary}
\end{table}

\begin{figure}[h]
\includegraphics [scale=0.5]{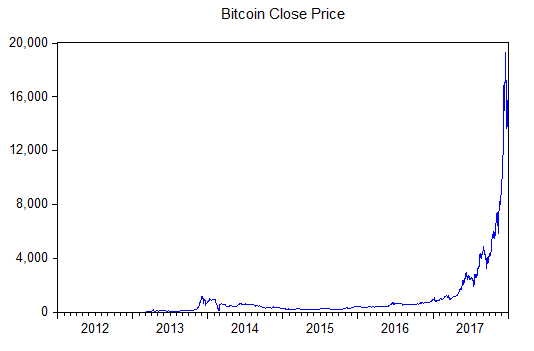} \includegraphics[scale=0.5]{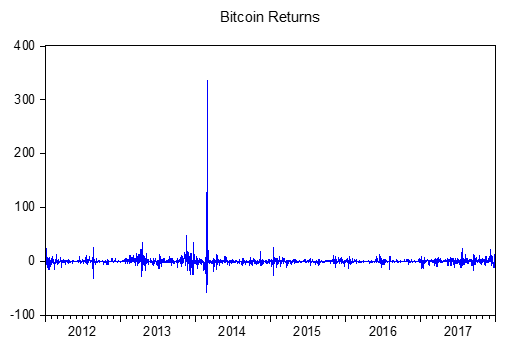}
\caption{Bitcoin Closing Price and Returns}
\label{fig:btc}
\end{figure}

Figure \ref{fig:btc} shows the bitcoin closing price and the first difference of closing price from 31 December 2011 to 31 December 2017.
Since the kurtosis for all cryptocurrencies under study is greater than 3, a non-Gaussian approach is necessary in order to model and understand the behaviour of cryptocurrencies’ returns. Jarque-Bera test rejects the null hypothesis that the returns are normally distributed at 5\% significance level. The purpose of this paper is to fit stable distribution to bitcoin returns and compare the goodness of fit with other heavy tailed distributions such as student-t distribution. Furthermore, we observe frequent jumps in the return data of bitcoin and consequently the assumption of finite variance in this case may not hold. Stable distribution which assumes infinite variance can explain the data better than the normal distribution in this case. To our knowledge, this is the first paper to use $\alpha-$stable process to model cryptocurrencies.

\section{Methodology: Stable Distributions}
 
Stable distributions are a rich class of distributions that includes the Gaussian, L{\`e}vy and Cauchy distributions in a family that allows for skewness and heavy tails, (Nolan, 1999b). Price changes are a result of new information into the market and of the re-evaluation of existing information, thus, changes in price represent the effect of many different bits of information (Fama,1965). Consequently, these bits of information may combine in additive fashion to produce stable distributions for daily, weekly or monthly periods. Stable distributions have been popular in statistical analysis of financial data since they are the only possible limiting distributions for sums of independent, identically distributed (i.i.d) random variables (Lux, 1996). Researchers have shown that changes in stock prices exhibit high volatility and statistical techniques such as the normal distribution which depend on the asymptotic theory of finite variance distributions are inadequte. An \(\alpha\)-stable is a L{\`e}vy process whose departure from the Brownian motion is controlled by the tail index \(\alpha\), which lies in the range \(0\textless\alpha\textless2\).  The additive property of the stable distribution can be expressed as follows: 
If \(X, X_1, X_2\cdots X_n\) are random variables, then for very positive integer \(n\), there exist constants \(a_n\textgreater0\),  \(B_n\) such that \begin{equation*}X_1+X_2+...+X_n \approx a_nX+B_n \end{equation*} thus LHS converges in distribution to the RHS. \footnote{For proofs and derivation of the stable distribution properties, see Samorodnitsky and Taqqu, 1994}

The most common parameterization for stable distribution is defined by Samorodnitsky and Taqqu (1994): A random variable $X$ is \(S(\alpha,\beta,\gamma,\delta)\) if it has  characteristic function. 

\[ E(\exp^{itX}) = \begin{cases} 
       \exp \biggl(-\gamma^\alpha | t|^\alpha \biggl[1-i\beta (\tan \frac{\pi\alpha}{2})(\text{sign} t)\biggl]+i\delta t \biggl), \alpha \neq 1 \\
      \exp\biggl(-\gamma|t| \biggl[1+i\beta \frac{2}{\pi}(\text{sign} t)\ln| t| \biggl]+i\delta t \biggl),  \alpha  = 1
   \end{cases}
\]   

The parameter \(\alpha\) is a measure of the thickness of the tails of the distribution and 
\begin{displaymath}
  \text{sign}t = \left\{
    \begin{array}{lr}
      1 & \text{ if  t \textgreater0}\\
      0 & \text{if     t =0}\\
     -1 & \text{if   t\textless0}\\
     \end{array}
   \right.
\end{displaymath} 

\begin{figure}
    \centering
    \includegraphics[scale=0.4] {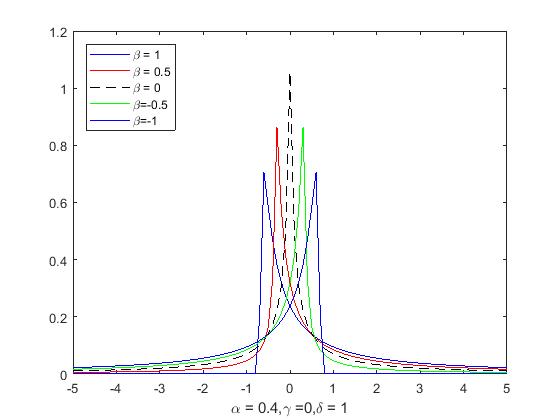} \includegraphics [scale=0.4]{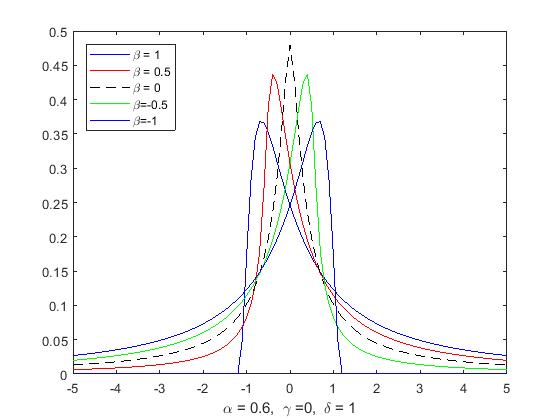}
    \includegraphics[scale=0.4]{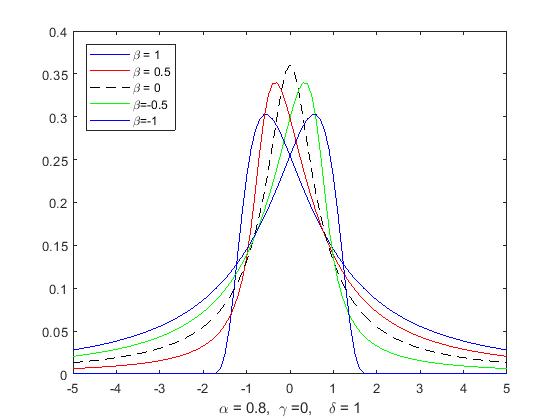} \includegraphics[scale=0.4]{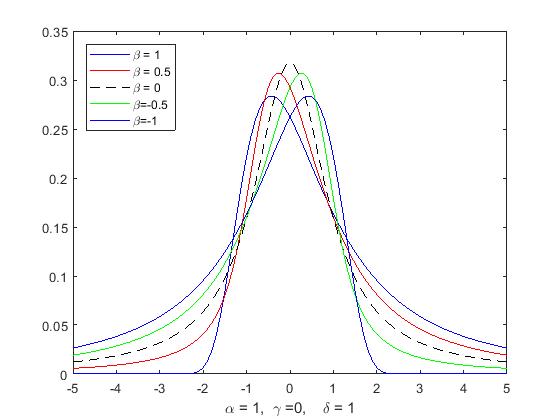}
     \includegraphics[scale=0.4] {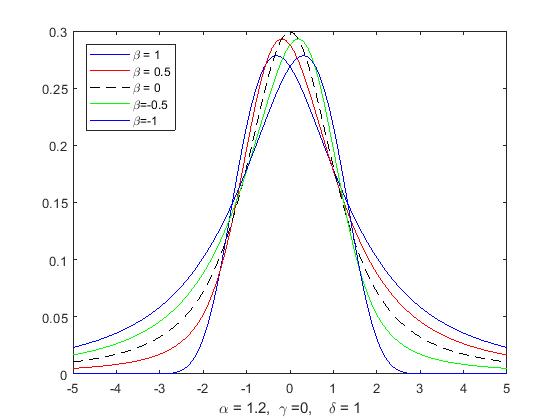} \includegraphics [scale=0.4]{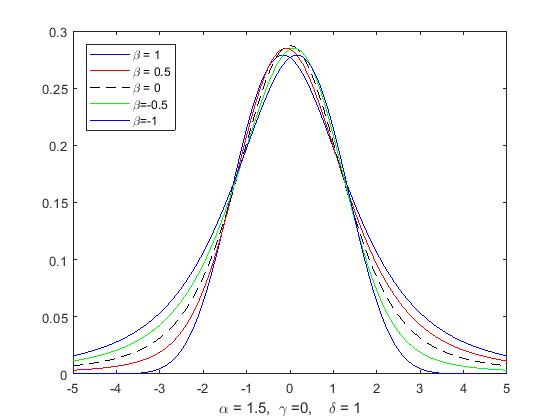}
     \includegraphics[scale=0.4]{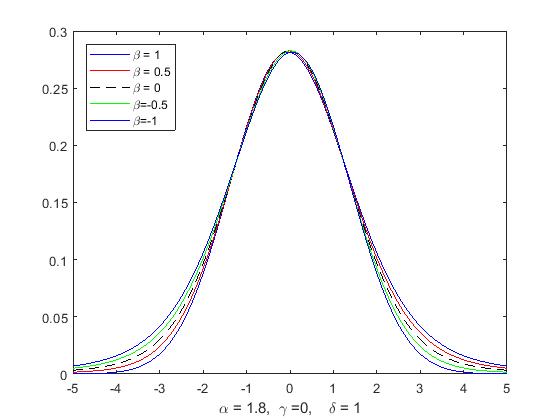} \includegraphics[scale=0.4]{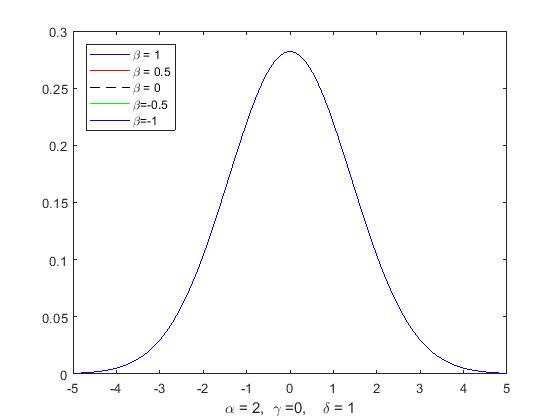}
    \caption{Shapes of $\alpha$-stable distribution for different values of  $\alpha$ and $\beta$}
    \label{fig:dist}
\end{figure}

A stable class has four parameters $\alpha, \beta, \gamma, \delta$, \footnote{0\(\textless \alpha \leq2\), \textbar\(\beta\)\textbar \(\leq\)min\((\alpha, 2-\alpha\)), \(\gamma \textgreater0\) -\(\infty \textless \delta \textless+ \infty\). When $ \alpha=2$, the resulting distribution is a normal distribution with mean $\delta$ and variance 2$\gamma^2$.} where   $\alpha$ describes the tail of the distribution,  $\beta$ is the skewness of the distribution, \footnote{When $\beta=0$, the distribution is symmetric, and if $\beta$ is greater than one the distribution is skewed to the right and if beta is less than one the distribution is skewed to the left.} $\delta$ is the location parameter, \footnote{$\delta$ is equal to the mean of the distribution if $\alpha$ equals one. $\delta$ shifts the distribution either to the left or to the right} and $\gamma$ is a scale parameter. \footnote{$\gamma$ compresses or extends the distribution about $\delta$ in proportion to $\gamma$}. As $\alpha$ increases, the effect of $\beta$ decreases. Figure \ref{fig:dist} shows the shapes of $\alpha$-stable distribution for different values of $\alpha$ and $\beta$.

When compared with other models used to capture leptokurtic features such as affine jump diffusion models and generalised hyperbolic models, \(\alpha\)-stable distribution is not only parsimonious with its four free parameters but also a creative model that is close to reality. Furthermore, setting \(\alpha\) below two effects a pure jump process with fat tails in the return distribution of cryptocurrencies and with such an infinite number of jumps, \(\alpha\)-stable distribution incorporates extreme market movements traditionally handled by diffusion processes. The major drawback of \(\alpha\)-stable distribution is that the density and distribution functions do not have closed form solutions except for a few members of the stable family.The distribution functions of stable distribution are known analytically under rare situations, that is, Cauchy distribution where \(\alpha\) =1 and Gaussian distributions where \(\alpha=2\) and the stable law of characteristic component or L{\`e}vy with \(\alpha = 1/2\). \\

However, empirical efforts have been made to alleviate this challenge. For instance, estimators for scale parameter and characteristic component were suggested by Fama and Roll (1971) who further provided probability tables of symmetric members of stable class with finite mean. They also suggested estimators for scale parameter and characteristic component; and examined goodness of fit test a robustness check for data analysis. Koutrouvelis (1980) used a regression-type method for estimating the four parameters of a stable distribution and found that the estimators were consistent and unbiased when analyzing large sample sizes. Paulson, Holcomb and Leitch (1975) improved the work of Fama and Roll (1971) by relaxing the hypothesis that stable distribution is symmetric, \(\beta =0\). When they allowed \(\beta\) to vary, the maximum absolute difference between the empirical and fitted distribution decreased significantly by 50\% when compared to Fama and Roll (1971) procedure which is applicable only to symmetric distributions.\\
Since the closed form probability density function for stable distribution is unknown except for a few members of the stable family, most of the conventional methods in mathematical statistics could not be used. The probability densities of $\alpha$-stable random variables exist and are continuous but, with a few exceptions, they are not known in closed form, (Zolotarev 1986b). These exceptions are:\\
\begin{enumerate}
\item The Gaussian distribution \(S_2\)\((\sigma,0,\mu \))=N\((\mu,2\sigma^2\)). A Gaussian distribution is a special case of stable distribution with \(\alpha\)=2, such that \(N(\mu,\sigma^2\))=\(S(2,0,\frac{\sigma}{\sqrt{2}}, \mu\)), where \(\mu\) is the mean  and \(\sigma\) is the standard deviation of the normal distribution. As noted earlier, when \(\beta\) = 2 there is no effect on stable distribution as the resulting distribution will be a normal distribution. The probability density function is given by
\begin{displaymath}
\frac{1}{2\sigma\sqrt{\pi}}\text{exp}^{-(x-\mu)^2/4\sigma^2}
\end{displaymath}

\item The Cauchy distribution. The Cauchy distribution is also another form of stable distribution with \(\alpha\)=1 and \(\beta\)=0, such that \(\text{Cauchy}(\delta,\gamma\))= \(S_1\)\((1,0,\gamma,\delta \)), where \(\gamma\) is the scale parameter and \(\delta\) is the location parameter of the Cauchy distribution. The probability density function is given by
\begin{displaymath}
\frac{\gamma}{\pi((x-\delta)^2+\gamma^2)}, -\infty\textless x\textless \infty
\end{displaymath}
If X \(\sim\)  \(S_1\)\((\gamma,0,0 \)), then for x\(\textgreater0\), its can be shown that P(X\(\leq x\))=\(\frac{1}{2}+\frac{1}{\pi}\arctan( \frac{x}{\gamma}\))
\item The L{\`e}vy distribution is also special case of stable distribution where \(\alpha\)=0.5 and \(\beta\)=1. In other words, L{\`e}vy\((\delta,\gamma)\)= \(S_{1/2}\)\((0.5,1,\gamma,\delta\)). The probability density function is given by
\begin{displaymath}
\sqrt{\frac{\gamma}{2\pi}}\frac{1}{(x-\delta)^{3/2}} \text{exp}\biggl[\frac{-\gamma}{2(x-\delta)}\biggl], \delta\textless x\textless\infty\\
\end{displaymath}

The pdf is concentrated on \(\delta,\infty\). If X \(\sim\)  \(S_{1/2}\) \((0.5, 1, \gamma\),\(\delta\)), then for \(x\textgreater\)0
\begin{displaymath}
P(X\leq x)=2\biggl(1-\phi\biggl(\sqrt{\frac{\gamma}{x}}\biggl)\biggl)
\end{displaymath}
where \(\phi\) denotes cumulative distribution function of the \(N(0,1)\) distribution.
\end{enumerate}

\section{Parameter Estimation}
DuMouchel (1971), (1973) was the first to propose the method of Maximum Likelihood (ML) principle to bracketed data in order to estimate parameters of stable distribution; and further provided a table of the asymptotic standard deviations and correlations of the ML estimators. McCulloch (1986) introduced the quantile-based method to estimate the four parameters of a stable distribution using five pre-determined sample quantiles with the aid of accompanying tables. ML method has received wide acceptance and use in approximating stable parameters for financial data. Mittnik et al (1999) implemented FFT-based Monte Carlo procedure to compare ML method with quantile-based method of McCulloch (1986) and found that the ML method is not only fast but also performs accurately compared to pdf calculations based on direct numerical integration. They also concluded that unlike ML estimator which can be easily modified to accommodate complicated extensions, the quantile-based method cannot be extended to complex estimation problems such as regressions that contain stable paretian disturbances; or ARMA and GARCH models that are driven by stable paretian innovations.\\

Furthermore, Nolan and Ojeda (2013) showed that ML outperforms OLS regression method and the performance of ML increases as the error distribution deviates from normality. Nolan (1998), (1999a) implemented the parameterization used by Zolotarev‘s (1986), Samorodnitsky and Taqqu (1994), McCulloch (1985) and DuMouchel (1971) (1973) into STABLE programme that can be used to give reliable computations of stable densities. Nolan (2001a) warned that stable distribution should be used to summarise the shape of the distribution and not to make statements about tail probabilities. In this paper we use STABLE programme\footnote{The program STABLE is available from J. P. Nolan's website: http://fs2.american.edu/jpnolan/www/stable/stable.html} to estimate the parameters and densities of the stable distribution.

Numerical methods such as McCulloch (1986) quantile-based method and ML estimators have been developed as a result of the absence of closed form solutions.
Let X=\((X_1,....,X_T\)) be a vector of T i.i.d stable Paretian random variables, and also \(x\) \(\sim\)  \(S_\alpha\)\((\alpha,\beta, \gamma, \delta \)). Defining \(\theta\)=\((\alpha,\beta, \gamma, \delta \)), Mittnik et al (1999) developed a ML algorithm and showed that the estimate of \(\theta\) can be obtained by maximising the log-likelihood function
\begin{displaymath}
\ell (\theta,x)={\sum}_{i=1}^{T} \text{log}f(x_i , \theta)
\end{displaymath}
with respect to the unknown parameter vector \(\theta\).
DuMouchel (1973) applied ML estimation to stable distribution inference and defined the likelihood function by
\begin{displaymath}
L(\theta)=\prod_{k=1}^{n} S_{\alpha, \beta} \biggl(\frac{X_k-\delta}{\gamma} \biggl) \biggl /\gamma
\end{displaymath}
where \(\theta\)=\((\alpha,\beta, \gamma, \delta \)) based on \(x=(x_1, ...x_n) \) for a sample size \(n\).

Another technique to estimate the parameters of a stable distribution is the quantile based approach introduced by McCulloch (1986). Using bitcoin as an example, we have 2 186 independent drawings \(X_i\),  from stable distribution \(S_\alpha\)\((\alpha, \beta, \gamma, \delta \)). We let \(X_p\) be the \(p-th\) population quantile such that  \(S_\alpha\) \((X_p, \alpha, \beta, \gamma, \delta \)) = \(p\). Given the above, we let \(\widehat{X_p}\) be the corresponding sample quantile with continuity correction . Thus, \(\widehat{X_p}\) is therefore a consistent estimator of \(X_p\). McCulloch defined the following
\begin{displaymath}
\upsilon_\alpha = \frac{X_{.95} - X_{.05}}{X_{.75}-X_{.25}};         \hspace{2cm}
\upsilon_\beta = \frac{X_{.95} + X_{.05} - 2X_{.5}}{X_{.95} - X_{.05}}
\end{displaymath}
where \(\upsilon_\alpha \) and \(\upsilon_\beta\) are independent of \(\gamma\) and \(\delta\). By letting \(\widehat{\upsilon_\alpha}\) and \(\widehat{\upsilon_\beta}\) be corresponding values of  \(\upsilon_\alpha \) and \(\upsilon_\beta\), respectively, and given that \(\upsilon_\alpha \) and \(\upsilon_\beta\) are functions of
\(\alpha\) and \(\beta\), the following relationship can be established:
\begin{displaymath}
\upsilon_\alpha = \phi_1(\alpha, \beta);                  \hspace{2cm}                           \upsilon_\beta =\phi_2(\alpha, \beta) \\
\end{displaymath}
The above relationship can further be inverted to yield the following
\begin{displaymath}
\alpha = \psi_1(\upsilon_\alpha,\beta);                 \hspace{2cm}                                               \beta = \psi_2( \upsilon_\alpha, \beta)  \\
\end{displaymath}
The parameters of \(\alpha\) and \(\beta\) may now be consistently estimated by
\begin{displaymath}
\hat{\alpha}=\psi_1(\widehat{\upsilon_\alpha}, \widehat{\upsilon_\beta});                   \hspace{2cm}                  \hat{\beta}=\psi_2(\widehat{\upsilon_\alpha}, \widehat{\upsilon_\beta});
\end{displaymath}

McCulloch (1986) showed the results of the relationship between  \(\upsilon_\alpha \) and \(\upsilon_\beta\) in table I-V of his paper. We also used quantile-based method to estimate the parameters of the stable distribution. A detailed approach of the sample characteristic method is found in Koutrouvelis (1980), Kogon and Williams (1998) and further clarified by Kateregga, Mataramvura and Zhang (2017).  In literature, ML Method has been found to yield consistent and accurate parameters of the stable distribution. Tables \ref{tab:btc_est}-\ref{tab:rpp_est} show results for different parameter estimation techniques for the three cryptocurrencies under study.
\begin{table}[ht!]
    \centering
    \begin{tabular}{lllll}
    \hline
    Estimation method & $\alpha$ & $\beta$ & $\gamma$ & $\delta$ \\ \hline
    Maximum Likelihood & 1.186$\pm$0.061	  & 0.111$\pm$0.086 & 1.564E-2$\pm$8.365E-4 & 2.515E-3 $\pm$1.057E-3\\ 
    Quantile-based method  & 1.1921	& -0.0301	& 0.015697	& 0.00241675\\ 
    Sample characteristic &1.3234	& 0.0539	& 0.0172434	 & 0.00333936\\ 
    \hline
    \end{tabular}
    \caption{Estimates of \(\alpha\)-stable distribution for Bitcoin}
    \label{tab:btc_est}
\end{table}

\begin{table}[ht!]
    \centering
    \begin{tabular}{lllll}
    \hline 
    Estimation method & $\alpha$ & $\beta$ & $\gamma$ & $\delta$ \\ \hline
    Estimation method & 1.186$\pm$0.061	  & 0.111$\pm$0.086 & 1.564E-2$\pm$8.365E-4 & 2.515E-3 $\pm$1.057E-3\\ 
    Quantile-based method  & 1.1921	& -0.0301	& 0.015697	& 0.00241675\\ 
    Sample characteristic &1.3234	& 0.0539	& 0.0172434	 & 0.00333936\\ 
    \hline
    \end{tabular}
    \caption{Estimates of \(\alpha\)-stable distribution for Etherium}
    \label{tab:eth_est}
\end{table}

\begin{table}[ht!]
    \centering
    \begin{tabular}{lllll}
    \hline
    Estimation method & $\alpha$ & $\beta$ & $\gamma$ & $\delta$ \\ \hline
    Maximum Likelihood & 1.1750$\pm$0.083 & 0.093$\pm$0.012 & 3.257E-2$\pm$2.407E-3 & 5.155E-3$\pm$2.30E-3\\ 
   Quantile-based method  & 1.1750	& 0.1321	&0.3166E-1	& -0.2868E-2  \\
  Sample characteristic & 1.2642	& 0.0621	&  0.3541E-1	 &  -0.5560E-2  \\ 
  \hline
    \end{tabular}
    \caption{Estimates of \(\alpha\)-stable distribution for Ripple}
    \label{tab:rpp_est}
\end{table}

\section{ Results and Goodness of fit test}
There are different ways that we can explore to establish whether the data is from a stable distribution. Nolan (1999), (2001) underscored that many heavy tailed distributions are not stable. We therefore need to test Nolan‘s (1999) proposition by comparing estimates from three different estimation techniques namely ML by DuMouchel (1971), quantile-based method by McCulloch (1986) and lastly the sample characteristic method proposed by Koutrovelis (1980), and Kogon and Wiliams (1998). For more details on the parameterizations of the characteristic function, refer to Zolotarev (1986). The argument is that these different methods are consistent estimators of parameters of a stable distribution. If the estimates are close then the hypothesis that the data is drawn from an $\alpha$-stable distribution is supported. However, Nolan (1999) does not state a boundary of how close the estimates should be relative to each other. \\ 

We use the non-parametric Kolgomorov-Smirnov test (K-S test) (Marsey, 1951)  to compare the goodness of fit for the three subclasses of stable distributions and the student-t distribution for each cryptocurrency under study. We further examine the estimation technique that yields the best fit for the cryptocurrencies under study. Most heavy tailed continuous distributions used in financial econometrics such as Log-logistic, Weibull and Log-normal only assume a positive vector of returns, hence, we could not use them. Although Generalised Gamma and Generalised Extreme Value distributions, among others, can be used with a vector of negative and positive values, they were insignificant at all levels and we included only L{\`e}vy distribution as an example of that case. We performed K-S test with estimates obtained using quantile-based and sample characteristic methods and we only include the maximum likelihood which was robust and yielded a higher $p-$value than other methods. \\

\begin{table}[ht!]
    \centering
    \begin{tabular}{lllll}
      \hline
 Sig. level &\(\alpha\)- Stable & Cauchy  & Student-t & L{\`e}vy\\ \hline
    20\% & 0.0229 & 0.0229 & 0.0229 & 0.0229 \\ 
    10\%  & 0.0261&0.0261 &0.0261&0.0261\\ 
     5\% &0.0290***& 0.0290 & 0.0290*** &0.0290\\ 
    1\%  & 0.0347***&0.0347*** &0.0347***& 0.0347 \\ 
test stat & 0.0261 & 0.0292 & 0.0270&0.2923\\ 
p-value & 0.0989  & 0.0472 &0.0805  & 3.0913E-163\\ 
    \hline
    \end{tabular}
    \caption{Results of the K-S test for Bitcoin}
    \label{tab:ks_btc}
\end{table}

\begin{table}[ht!]
    \centering
    \begin{tabular}{lllll}
    \hline
        Sig. level & \(\alpha\)- Stable & Cauchy  & Student-t & L{\`e}vy\\ \hline
    20\% & 0.0343*** & 0.0343 & 0.0343 & 0.0343 \\ 
    10\%  & 0.0392***&0.0392 &0.0392***&0.0392\\ 
     5\% &0.0435***& 0.0435& 0.0435*** &0.0435\\ 
    1\%  & 0.0521***&0.0521*** &0.0521***& 0.0521 \\ 
test stat & 0.0337 & 0.0517 & 0.0371&0.2955\\
p-value & 0.2176  & 0.0110 &0.1350 & 2.7866E-74\\ 
\hline
    \end{tabular}
    \caption{Results of the K-S test for Etherium}
    \label{tab:ks_eth}
\end{table}

\begin{table}[ht!]
    \centering
    \begin{tabular}{lllll}
    \hline
      Sig. level & \(\alpha\)- Stable & Cauchy  & Student-t & L{\`e}vy\\ \hline
    20\% & 0.0313*** & 0.0313 & 0.0313 & 0.0313 \\ 
    10\%  & 0.0357***&0.0357 &0.0357***&0.0357\\ 
     5\% &0.0396***& 0.0396***& 0.0396*** &0.0396\\ 
    1\%  & 0.0475***&0.0475*** &0.0475***& 0.0475 \\ 
test stat & 0.0291 & 0.0394 & 0.0322&0.2864\\ 
p-value & 0.2725  & 0.0524&0.1746 & 6.5926E-84\\
\hline
    \end{tabular}
    \caption{Results of the K-S test for Ripple}
    \label{tab:ks_rpp}
\end{table}
* means that the p-value is higher than the corresponding significance level, hence we accept the null hypothesis that the data comes from the stated distribution \\

We note that $\alpha$-stable, Student-t and Cauchy distributions are significant at different levels for the cryptocurrencies under study, however, when comparing the p-values, we find more evidence in support of the $\alpha$-stable distribution than Student-t and Cauchy distributions since the p-value of alpha stable is higher than that of other distributions. We also find that Student-t distribution outperforms the Cauchy distribution when considering the p-values.  L{\`e}vy distribution was not significant even at 1\% level and this suggests that the distribution cannot be used to model cryptocurrencies and other highly speculative assets with similar characteristics.

\section{Conclusion}
In this paper we have applied $\alpha$-stable distribution to model cryptocurrency return data and compared the goodness of fit with other heavy tailed distributions used in financial econometrics. The empirical study shows that $\alpha$-stable distribution with parameters estimated by ML method is better fitted to model highly speculative cryptocurrencies particularly bitcoin, ethereum and ripple. The leptokurtic features that exist in bitcoin due to high volatility can be captured by an $\alpha$-stable distribution. For cryptocurrency data, we found that student-t distribution outperforms Cauchy distribution. However, the tail behavior of the data deviates from that of Stable Paretian distribution, a phenomenon that could be associated with a generalized Pareto or simple Pareto tail-index estimate above 2 which has been frequently cited as evidence against infinite-variance stable distributions. In a critique, McCulloch (1995) argued that the inference is invalid since a tail index above 2 can result from a stable distribution with $\alpha$ as low as 1.65. Future research can focus on incorporating $\alpha$-stable process in the price prediction of crytocurrencies and risk analysis in portfolio management.

\end{document}